\documentclass[prapplied,amsmath,reprint,floatfix,a4paper,superscriptaddress]{revtex4-1}

\usepackage[T1]{fontenc}
\usepackage{etex}
\usepackage{geometry}
\usepackage{color,tabularx,colortbl}

\usepackage[colorlinks = true,
            linkcolor = blue,
            urlcolor  = blue,
            citecolor = blue,
            anchorcolor = blue]{hyperref}
\usepackage{setspace}
\usepackage{pgfplots}
\usepackage{booktabs}
\usepackage{here}
\usepackage{multirow}
\usepackage{tikz, tikz-3dplot}
\tikzset{>=latex}
\usepackage[percent]{overpic}
\pgfplotsset{compat=1.13}
\usepgfplotslibrary{external} 
\usepgfplotslibrary{fillbetween}
\usepgfplotslibrary{colorbrewer}
\usetikzlibrary{pgfplots.groupplots,arrows,decorations.markings}
\usetikzlibrary{shapes.geometric,calc,patterns}
\usepackage[strict]{changepage}

\pgfmathsetmacro{\kJToMerg}{0.01}
\pgfmathsetmacro{\TtokOe}{10}
\pgfmathsetmacro{\TtokEmu}{10/(4*pi)}
\pgfmathsetmacro{\JtokErg}{1}

\definecolor{uni_blue}{RGB}{0,99,166}
\definecolor{uni_grey}{RGB}{102,102,102}
\definecolor{uni_red}{RGB}{167,28,73}
\definecolor{uni_orange}{RGB}{221,72,20}
\definecolor{uni_yellow}{RGB}{246,168,0}
\definecolor{uni_light_green}{RGB}{148,193,84}
\definecolor{uni_mint}{RGB}{17,137,122}

\pgfplotsset{
colormap={myColMap}{[5pt] 
	    color(0pt)=(white);
        color(10pt)=(uni_blue!10);
        color(20pt)=(uni_blue!20);
        color(30pt)=(uni_blue!30);
        color(40pt)=(uni_blue!40);
        color(50pt)=(uni_blue!50);
        color(60pt)=(uni_blue!60);
        color(70pt)=(uni_blue!70);
        color(80pt)=(uni_blue!80);
        color(90pt)=(uni_blue!90);
        color(100pt)=(uni_blue);
	    }
}
\pgfplotsset{
colormap={myColMapReduced}{[5pt] 
	    color(0pt)=(white);
	    color(500pt)=(white);
	    color(505pt)=(uni_blue);
        color(1000pt)=(uni_blue);
	    }
}

\pgfplotscreateplotcyclelist{uni}{
    {color=uni_blue,mark=none},
    {dashed,color=uni_yellow,mark=none},
    {color=uni_red,mark=none},
    {dashed,color=uni_light_green,mark=none},
    {color=uni_orange,mark=none},
    {dashed,color=uni_grey,mark=none},
    {color=uni_light_green,mark=none},
    }

\begin{document}

\title{Spin-canting effects in GMR sensors with wide dynamic field range} 

\author{Clemens Muehlenhoff}
\email{clemens.muehlenhoff@physik.uni-augsburg.de}
\affiliation{Institute of Physics, University of Augsburg, Augsburg 86159, Germany}
\affiliation{Infineon Technologies AG, 85579 Neubiberg, Germany}
\author{Christoph Vogler}
\email{christoph.vogler@univie.ac.at}
\affiliation{Christian Doppler Laboratory for Advanced Magnetic Sensing and Materials, Faculty of Physics, University of Vienna, Boltzmanngasse 5, 1090 Vienna, Austria}
\author{Wolfgang Raberg}
\affiliation{Infineon Technologies AG, 85579 Neubiberg, Germany}
\author{Dieter Suess}
\affiliation{Christian Doppler Laboratory for Advanced Magnetic Sensing and Materials, Faculty of Physics, University of Vienna, Boltzmanngasse 5, 1090 Vienna, Austria}
\author{Manfred Albrecht}
\affiliation{Institute of Physics, University of Augsburg, Augsburg 86159, Germany}

\begin{abstract}
Magnetoresistive (xMR) sensors find extensive application in science and industry, replacing Hall sensors in various low field environments. While there have been some efforts in increasing the dynamic field range of xMR sensors, Hall sensors remain to dominate high field applications due to their wide linear range.
Using a perpendicular magnetized reference system and an in-plane free layer allows us to overcome this disadvantage of xMR sensors, and, furthermore, investigate spin-canting effects in interlayer exchange coupled perpendicular synthetic antiferromagnets (p-SAF). We created p-SAFs with exchange coupling fields of up to 10\,kOe, based on magnetic Co/Pt multilayer systems.  The p-SAFs are either designed as "single" p-SAFs, where two Co/Pt multilayers are interlayer exchange coupled via a 4\,$\text{\AA}$ thick Ru spacer, or as "double" p-SAFs, where an additional Co layer is interlayer exchange coupled to the top multilayer. These p-SAFs are used for giant magnetoresistance (GMR) sensors with wide dynamic field range. By using a p-SAF as the reference system and employing an in-plane magnetic layer as the GMR's free layer, the linear range can be effectively increased limited only by the p-SAF's switching fields. Additionally, the magnetic anisotropy of the in-plane free layer is fully controlled, which allows saturation fields by design. With this, the entire spectrum from parallel to antiparallel alignment of free and reference layer is exploited, which yields the full GMR signal potential. Different configurations were investigated, ranging from free layer magnetic saturation at lower to far higher fields than the p-SAF's switching fields. We can show through micromagnetic simulations that certain GMR transfer curves are dominated by spin-canting effects in the interlayer exchange coupled reference system. Finally, our simulation results lay out the correlation of the p-SAF's design parameters and its magnetization reversal behavior.
\end{abstract}
\maketitle
\section{INTRODUCTION}
Magnetoresistive (xMR) effects, such as the giant (GMR) \cite{Butler1995} or the tunneling magnetoresistance (TMR) \cite{Zhang2015}, are already established devices for modern day technologies \cite{Hirohata2014, Hirohata2015, Zheng2019, Tsymbal2016}. Besides their application in Hard Disk Drive (HDD) read heads, xMR sensors have also replaced Hall sensors in many areas, due to their superior sensitivity, high frequency bandwidth, and low power consumption \cite{Ikeda2008, Binasch1989}.
However, there are still properties of the Hall sensor, in which xMR based sensors cannot compete. These properties include an out-of-plane sensitivity, as well as a linear field range that is orders of magnitude higher than that of xMR sensors. Within the automotive industry, a wide dynamic field range is a requirement for current sensing applications, where currents may be up to $1\,\text{kA}$ large, thus generating strong magnetic fields. It is therefore estimated that an applicable magnetic sensor will need a dynamic field range of more than $\pm2$\,kOe \cite{Ogasawara2018}.
One successful method to increase the field range of TMR sensors is the use of a magnetic vortex structure in the magnetic free layer (FL), which yields linear ranges in the order of 500\,Oe \cite{Suess2018, PhysRevApplied.10.054056, Wurft2017, Wurft2019}. In this work, we utilize perpendicular magnetic anisotropy (PMA) in the reference system, while the FL has in-plane anisotropy \cite{Mancoff2000,Ogasawara2019,Nakano2017a,Wei2009,Wisniowski2012,Matthes2015b, Matthes2015, Nakano2018a}. This cross-geometric anisotropy in xMR sensor technology has realized TMR sensors of up to $\pm$5.6\,kOe dynamic range \cite{Ogasawara2019}. Here, we present GMR sensors with perpendicular synthetic antiferromagnets (p-SAF) as reference system of up to 10\,kOe exchange coupling field $H_\text{ex}$ \cite{Yakushiji2015, Yakushiji2017a}, enabling accordingly wide dynamic ranges. We study the impact of the FL saturation field on GMR transfer curves, using superconducting quantum interference device - vibrating sample magnetometry (SQUID-VSM), polar magneto-optic Kerr effect (MOKE), and GMR measurement setups. The high $H_\text{ex}$ in the reference system and the FL saturation field control are achieved by using Co/Pt multilayer (ML) systems \cite{Johnson1996, Yakushiji2015, Chatterjee2014, Yakushiji2010, Chatterjee2018}, where PMA stems from strong spin-orbit coupling and \textit{d}-\textit{d} orbital hybridization at the interface \cite{Johnson1995, Johnson1996, Nakajima1998}. The behavior in magnetization reversal and magnetoresistance are further investigated through micromagnetic simulations, focusing on switching fields and spin-canting effects \cite{Ackermann2018, Fernandez-Pacheco2019} within the p-SAF reference system.
  
\section{EXPERIMENTAL DETAILS}
\begin{figure}
\includegraphics{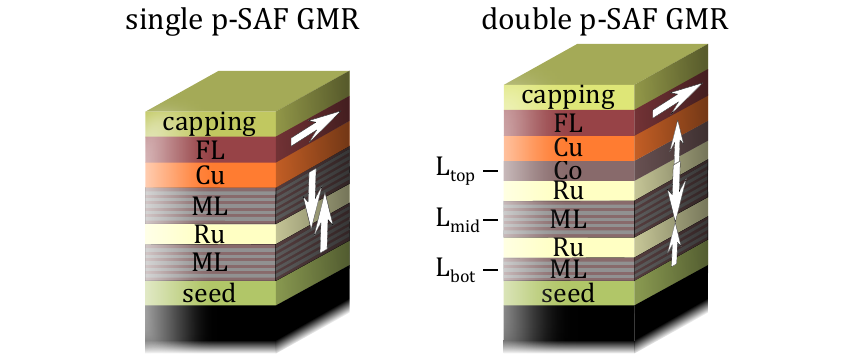}%
 \caption{\small\label{fig:pSAF-illustrations}Illustrations of the employed single and double p-SAF GMR stacks. While the single p-SAF uses only one Ru spacer layer, the double p-SAF incorporates two spacer layers for interlayer exchange coupling. Thus, double p-SAFs consist of three antiferromagnetically coupled ferromagnetic layers, L$_\text{bot}$, L$_\text{mid}$, and L$_\text{top}$. White arrows illustrate magnetization directions at zero field.}
\end{figure}

Films were deposited onto thermally oxidized silicon (100) substrates in a BESTEC magnetron-sputtering chamber with a base pressure of 1$\times$10$^{-8}$\,mbar. Argon was used as a sputter gas during the deposition process with a pressure of 5$\times$10$^{-3}$\,mbar at room temperature. 
For this work, two different types of p-SAF reference systems were fabricated, which differ in their number of interlayer exchange coupling spacer layers \cite{Bruno1995}. For single p-SAF systems, we used the following stack structure: substrate/{\allowbreak}Ta(50)/{\allowbreak}Ru(100)/{\allowbreak}Pt(20)/{\allowbreak}[Co(3)/{\allowbreak}Pt(2)]$_5$/{\allowbreak}Co(3)/{\allowbreak}Ru(4)/{\allowbreak}[Co(3)/{\allowbreak}Pt(3)]$_5$/{\allowbreak}Co(4)/{\allowbreak}cap (thicknesses in $\text{\AA}$). The double p-SAF structures are composed of substrate/{\allowbreak}Ta(50)/{\allowbreak}Ru(100)/{\allowbreak}Pt(20)/{\allowbreak}[Co(3)/Pt(2)]$_{\text{X}}$/{\allowbreak}Co(3)/{\allowbreak}Ru(4)/{\allowbreak}[Co(3)/Pt(3)]$_{\text{Y}}$/{\allowbreak}Co(3)/{\allowbreak}Ru(4)/{\allowbreak}Co(10)/{\allowbreak}cap, where X and Y refer to the respective number of bilayers in the ML, which vary depending on the sample's purpose. P-SAFs reveal the highest exchange fields for X=4 and Y=7, whereas p-SAFs for the investigation of FL saturation fields and spin-canting effects are built with X=6, and Y=8. Illustrations of the two types of samples are given in Fig.~\ref{fig:pSAF-illustrations}, showing a single and double p-SAF GMR stack with three ferromagnetic layers L$_\text{bot}$, L$_\text{mid}$, and L$_\text{top}$ for the double p-SAF. 
For complete GMR sensor structures, a 20\,$\text{\AA}$ Cu spacer layer, followed by a magnetic FL were deposited on top of L$_\text{top}$. Four different FL systems S$_\text{x}$ were fabricated: \\
S$_\text{a}$: Co(30), \\
S$_\text{b}$: [Co($t_{\text{Co}}$)/Pt(2.5)]$_3$, \\
S$_\text{c}$: Co(30)/Pt(2.5)/[Co(3.5)/Pt(2.5)]$_3$, and \\
S$_\text{d}$: Co(30)/CoFe(30). \\
As a capping layer, Pt(20)/Ta(50) was used. The out-of-plane magnetization of all samples was measured in a SQUID-VSM magnetometer and a polar MOKE setup at room temperature. Furthermore, GMR measurements were conducted at room temperature in Van-der-Pauw configuration.

\section{EXPERIMENTAL RESULTS AND DISCUSSION}
\begin{figure}
\includegraphics{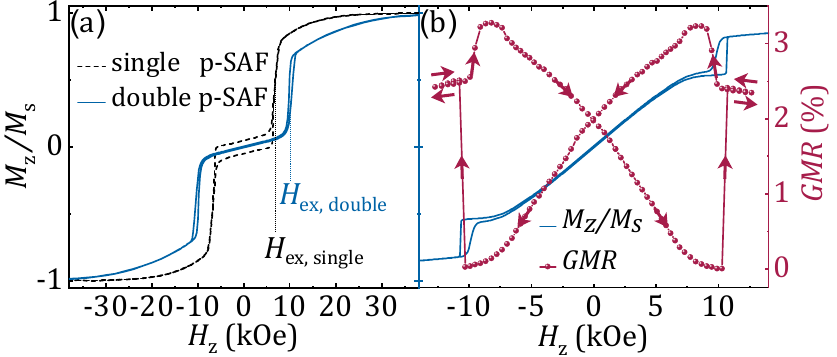}%
 \caption{\small\label{fig:extremeWideLR}(a) Comparison of normalized magnetization curves, taken from SQUID-VSM measurements of a single and double p-SAF. (b) Magnetization and GMR signal of a GMR sensor with a double p-SAF reference system, yielding $H_{\text{ex}}$ = 10\,kOe, and a FL system of type S$_\text{b}$ ($t_\text{Co}$=20\,$\text{\AA}$). Red arrows in the GMR transfer curve represent the direction of the magnetic field sweep.}
\end{figure}
\begin{figure}
\includegraphics{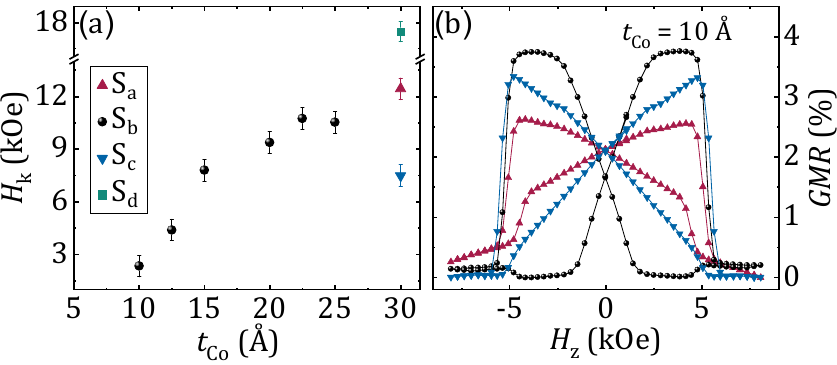}%
 \caption{\small \label{fig:Hkcontrol}(a) Free layer $H_{\text{k}}$ values for different FL systems and FL Co layer thicknesses with (b) corresponding GMR transfer curves of single p-SAF GMR sensors.}
\end{figure}
Figure~\ref{fig:extremeWideLR} shows the normalized magnetizations, $M_\text{z}/M_\text{S}$, for a single and a double p-SAF in out-of-plane fields, here defined as $H_{\text{z}}$ pointing in z-direction. Even at $H_\text{z}<H_\text{ex}$ a change in $M_\text{z}/M_\text{S}$ of $0.9\frac{\%}{\text{kOe}}$ is detected for both types of p-SAFs, indicating the occurrence of small spin-canting throughout the field range. Double p-SAF structures yield higher $H_{\text{ex}}$, as in this case $H_{\text{ex}}$~=~10\,kOe. However, the magnetization still increases at fields above $H_{\text{ex}}$, especially for double p-SAF structures, which again suggests spin-canting of the p-SAF reference system. Systems with $H_{\text{ex}}$ of up to 10\,kOe are also present in GMR structures using a FL system of type S$_\text{b}$ with $t_\text{Co}$=20\,$\text{\AA}$, as shown in Fig.~\ref{fig:extremeWideLR}(b). The resistivity $R$, here represented as the GMR effect with 
\begin{equation}
\textit{GMR} \text{(\%)}=100 \times \frac{R(H_{\text{z}})-R_{\text{min}}}{R_{\text{min}}},
\label{eq:GMR}
\end{equation}
shows abrupt changes at the fields of p-SAF layer switching. The FL is designed to magnetically saturate around 9\,kOe, which can be seen in the magnetization curves (Fig.~\ref{fig:extremeWideLR}(b)). However, similarly to Fig.~\ref{fig:extremeWideLR}(a), the magnetization of the GMR system still increases slightly above $H_\text{ex}$. Even more unexpectedly, the resistivity at fields above 10\,kOe is still higher than at zero field. This indicates that the p-SAF is not fully saturated and is likely to have its top layer magnetization in a near antiparallel state with that of the saturated FL in order to induce a GMR effect.\\
\begin{figure*}
\centering
\includegraphics{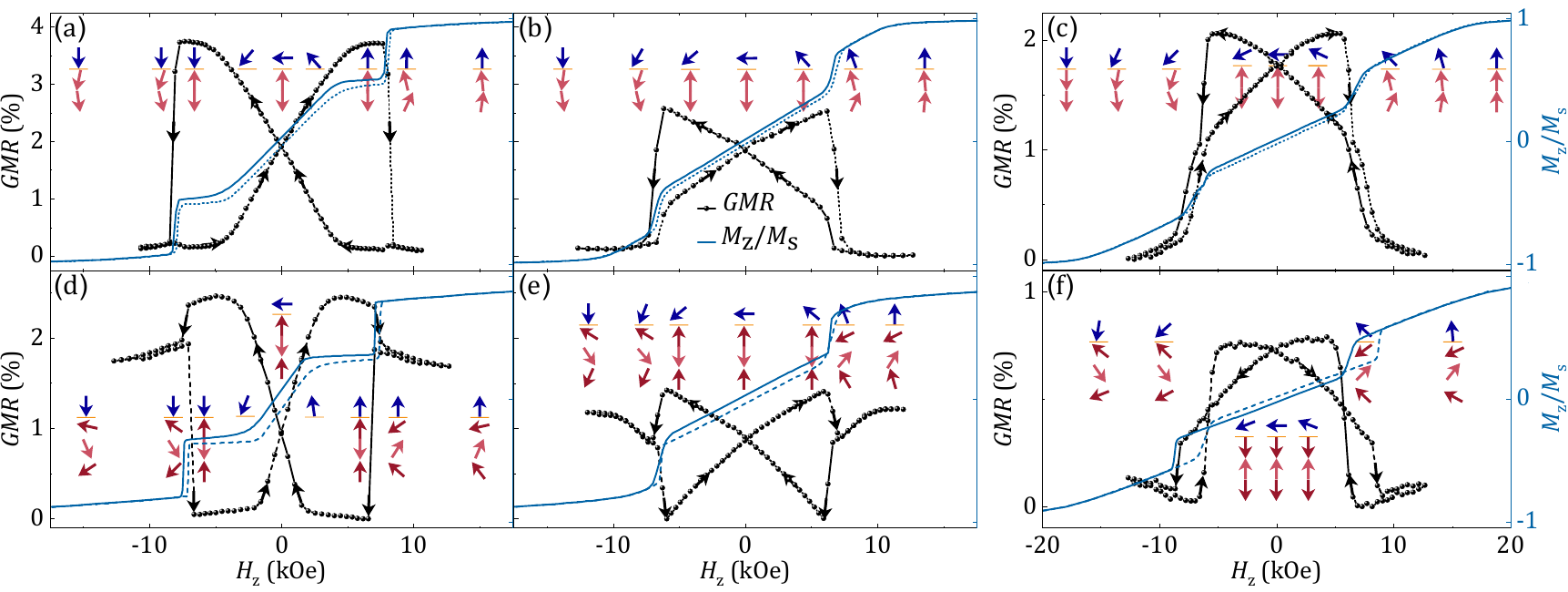}%
 \caption{\small GMR signals and normalized magnetizations for GMR sensors in out-of-plane fields. GMR sensors with (a-c) single and (d-f) double p-SAF reference system (X=6, Y=8), and free layer (a,c)~$H_{\text{k}}\leq H_{\text{ex}}$, (b,e)~$H_{\text{k}}> H_{\text{ex}}$, and (c,f)~$H_{\text{k}}\gg H_{\text{ex}}$. The respective FL systems are (a) S$_\text{b}$ ($t_\text{Co}$=12.5\,$\text{\AA}$), (b) S$_\text{b}$ ($t_\text{Co}$=25.0\,$\text{\AA}$), (c) S$_\text{d}$, (d) S$_\text{b}$ ($t_\text{Co}$=10.0\,$\text{\AA}$), (e) S$_\text{c}$, and (f) S$_\text{d}$. The arrows illustrate possible magnetization directions of the respective layers, with blue representing the free layer and red representing the layers of the p-SAF reference system. Solid lines represent a field sweep from positive to negative saturation, dashes lines from negative to positive saturation.\label{fig:GMRconfigs}}
\end{figure*} 
In order to  exploit the full spectrum of parallel to antiparallel alignment of free and reference layer, the  in-plane FL anisotropy needs to be adjusted in such a way that the FL saturates before or at the p-SAF switching field of the reference system. For TMR structures with an MgO barrier and a CoFeB free layer this can be achieved by choosing CoFeB to be appropriately thin so that orbital hybridization at the MgO/CoFeB interface adds perpendicular magnetic anisotropy to the FL system \cite{Wisniowski2012, Nakano2017a}. For our GMR sensors, however, we use four different FL systems to change the FL's magnetic anisotropy. In Fig.~\ref{fig:Hkcontrol}(a) the effective anisotropy field $H_{\text{k}}$ of each FL system is given, taken from out-of-plane SQUID-VSM measurements including the standard deviation from multiple fabrication cycles used as error bars. Figure~\ref{fig:Hkcontrol}(b) shows examples of corresponding GMR transfer curves. The simplest FL system S$_\text{a}$, represented as red upwards triangles, saturates at 12.5\,kOe in $H_\text{z}$. Is the FL designed as a multilayer system of type S$_\text{b}$, the saturation field can be easily adjusted via the specific Co thickness $t_\text{Co}$ in the ML, here shown as black dots. The effective magnetic anisotropy favors an in-plane easy axis magnetization for $t_\text{Co}\geq10$\,$\text{\AA}$ \cite{DenBroeder1991, Yakushiji2015b}, due to the large contribution of magnetic shape anisotropy. For growing  $t_\text{Co}$ the magnetic shape anisotropy increases, leading to a FL saturation at higher out-of-plane fields. Similarly, the FL magnetic saturation field is decreased to approximately 7.5\,kOe for a sample with a type S$_\text{c}$ FL, where a layer of Co(30) with strong in-plane magnetic anisotropy is coupled to a ML with PMA, here represented with blue downwards triangles. Lastly, adding a CoFe(30) layer to the Co layer in the S$_\text{d}$ FL, shown as teal rectangle in Fig.~\ref{fig:Hkcontrol}(a), significantly increases $H_{\text{k}}$ to 17.5\,kOe, which is partly due to the CoFe layer acting as a spacer between the Co free and the Pt capping layer, thus preventing PMA at the Co/Pt interface. The corresponding GMR transfer curves are shown later in Figs.~\ref{fig:GMRconfigs}(c) and \ref{fig:GMRconfigs}(f).

Using the structures mentioned above, we systematically changed $H_{\text{k}}$ of the FL for single and double p-SAF GMR systems, as shown in Fig.~\ref{fig:GMRconfigs}, with single p-SAFs in Figs.~\ref{fig:GMRconfigs}(a-c) and double p-SAFs in Figs.~\ref{fig:GMRconfigs}(d-f). GMR transfer curves as well as normalized $M$-$H_\text{z}$ hysteresis loops from a SQUID-VSM are plotted. The arrows illustrate the assumed magnetization direction of the respective magnetic layers (from positive saturation to negative). Note the different scales for Figs.~\ref{fig:GMRconfigs}(c)~and~\ref{fig:GMRconfigs}(f). From our measurements, 3 distinct configurations regarding the relation of FL $H_{\text{k}}$ to p-SAF $H_{\text{ex}}$ are visible for both single and double p-SAF GMRs in order of appearance in Fig.~\ref{fig:GMRconfigs}: (a,d)~$H_{\text{k}}\leq H_{\text{ex}}$, (b,e)~$H_{\text{k}}> H_{\text{ex}}$, and (c,f)~$H_{\text{k}}\gg H_{\text{ex}}$. 
Generally, the decrease in sensor sensitivity, i.e. the change of GMR over field, with increasing $H_{\text{k}}$ is well visible, as expected. In the following, we discuss the results in more detail, and argue that spin-canting effects offer an explanation for the observed magnetization and GMR transfer curves.\\
The single p-SAF GMR sensors in Figs.~\ref{fig:GMRconfigs}(b) and \ref{fig:GMRconfigs}(c) show their smallest resistance for $H_\text{z}> H_\text{ex}$, when the FL saturation is reached at $H_\text{k}$. Spin-canting effects in the p-SAF system aren't visible, due to the change of magnetization in the FL. The exception is the sample of Fig.~\ref{fig:GMRconfigs}(a) with $H_\text{k}<H_\text{ex}$. Here, a small increase in $R$ is visible at $H_\text{ex}$, followed by a decrease in $R$ for $H_\text{z}>H_{\text{ex}}$. This indicates very small spin-canting of the p-SAF reference system for $H_\text{z}>H_\text{ex}$. The double p-SAF GMR sensors in Figs.~\ref{fig:GMRconfigs}(d-f), on the other hand, show an unexpected high $R$ at $H_\text{z}>H_{\text{ex}}$. The resistance in Fig.~\ref{fig:GMRconfigs}(d) gradually decreases for increasing $H_\text{z}>H_\text{ex}$. Since the FL of the sample saturates already before $H_\text{ex}$, this decrease is caused by a spin-canted state of the p-SAF reference system and its gradual magnetic alignment with the FL magnetization at higher fields. In Fig.~\ref{fig:GMRconfigs}(e), the resistance further increases for increasing $H_\text{z}>H_\text{ex}$, until the FL saturation magnetization is reached. A different GMR behavior can be seen in Fig.~\ref{fig:GMRconfigs}(f), where $H_\text{k}\gg H_\text{ex}$. Here, $R$ is smaller for $H_\text{z}>H_\text{ex}$, indicating a more parallel configuration of FL and reference layer magnetization direction than for $H_\text{z}<H_\text{ex}$. However, with further increasing $H_\text{z}>H_\text{ex}$, magnetization directions rotate to an increasingly antiparallel alignment, owed to the strong spin-canting of the p-SAF reference system. This leads an increase in $R$ for high fields, in contrast to the decrease that is observed in the case of the same FL but with a single p-SAF reference system, as shown in Fig.~\ref{fig:GMRconfigs}(c).\\
Summarized, we observe only a small spin-canting behavior for single p-SAF GMR sensors, but strong spin-canting for double p-SAF GMR samples at fields above $H_\text{ex}$, which yields significantly different GMR transfer curves.\\
\\
\begin{figure}
\includegraphics{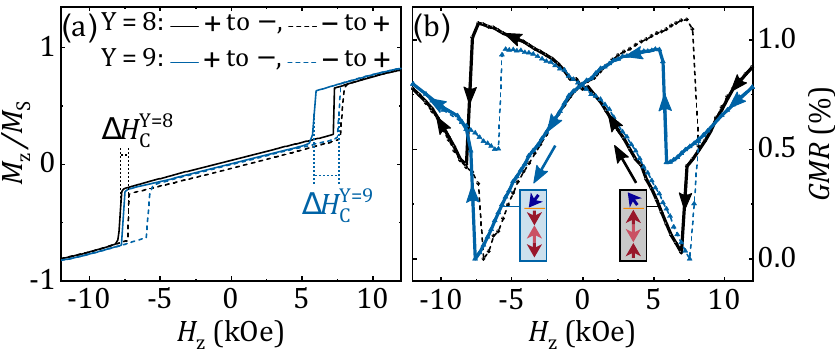}%
 \caption{\small \label{fig:pSAFs-Switching}(a) SQUID-VSM and (b) GMR measurements of double p-SAF GMR sensors with differing bilayer number Y=8 (black) and Y=9 (blue) in L$_\text{mid}$. Positive to negative field sweeps are drawn in solid lines with arrows to guide the eye, negative to positive field sweeps are drawn in dashed lines.}
\end{figure}
Depending on the magnetic moments within the double p-SAF structures, either L$_\text{mid}$ or the outer layers, L$_\text{bot}$ and  L$_\text{top}$, switch first. The difference is shown in Fig.~\ref{fig:pSAFs-Switching}, where two consecutively fabricated double p-SAF GMR sensors yield a different switching behavior due to a change of $M_\text{S}t_\text{ML}$ in L$_\text{mid}$ by design, with $t_\text{ML}$ being the total film thickness of the ML. This increase of $M_\text{S}t_\text{ML}$ is achieved by using a higher number of Co/Pt bilayers Y=9 instead of Y=8 \cite{Yakushiji2010}. In Fig.~\ref{fig:pSAFs-Switching}(a) it is visible that the total magnetization reversal and the underlying magnetization profile is similar for both samples, however, with the difference of the Y=9 samples showing a significantly higher hysteresis width $\Delta H_\text{C}$, which we define by the field difference of p-SAF switching fields $H_\text{C,1}$ and $H_\text{C,2}$. A further difference emerges in the comparison of GMR transfer curves in Fig.~\ref{fig:pSAFs-Switching}(b). As measurements from positive to negative fields are drawn in solid lines, whereas negative to positive field sweeps are drawn in dashed lines, the transfer curves of the two samples reveal quite different switching behaviors. 
\begin{figure}
\includegraphics{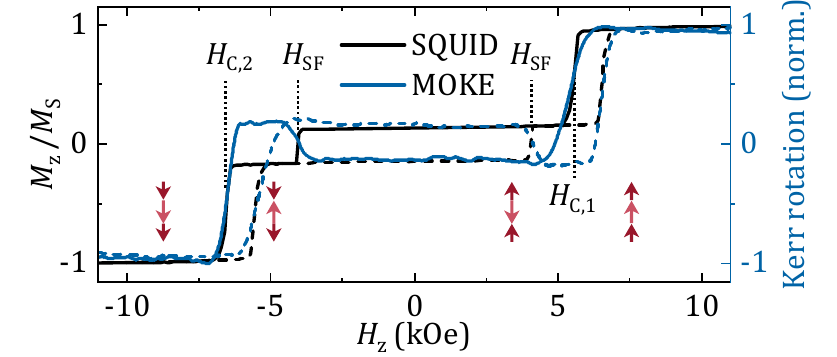}%
 \caption{\small\label{fig:pSAFs-SpinFlip}SQUID-VSM and MOKE measurement of an imbalanced p-SAF with occuring spin flip field $H_\text{SF}$.}
\end{figure}
In a sweep from positive to negative fields, the Y=8 sample of lower L$_\text{mid}$ magnetization jumps from a spin-canted state into a configuration of low GMR, with L$_\text{top}$ and L$_\text{bot}$ dominating the p-SAF, pointing parallel to the external magnetic field, and L$_\text{mid}$ pointing antiparallel (see illustration of magnetization in the right inset). Towards smaller and eventually negative fields, the GMR increases as the FL magnetization rotates into an antiparallel alignment with L$_\text{top}$. For the Y=9 sample the opposite p-SAF behavior occurs, with L$_\text{mid}$ dominating the p-SAF, leading to a decrease of GMR with decreasing fields, as illustrated in the left inset. Since our samples in Fig.~\ref{fig:GMRconfigs} were designed with balanced magnetic moments of the respective magnetic layers, even small fluctuations during the fabrication process may lead to a different switching behavior. This is demonstrated in the GMR sensor plotted in Fig.~\ref{fig:GMRconfigs}(f) in contrast to those plotted in \ref{fig:GMRconfigs}(d) and \ref{fig:GMRconfigs}(e).\\
It has to be noted that in case of highly imbalanced p-SAF structures, an additional switching can occur \cite{Yakushiji2015}, as presented in Fig.~\ref{fig:pSAFs-SpinFlip}. Here, SQUID-VSM and MOKE measurements show a spin flip of the p-SAF at $H_\text{SF}$ between the switching fields $H_\text{C,1}$ and $H_\text{C,2}$, where the coupled p-SAF system reverses its magnetization orientations, keeping the preferred antiparallel alignment. This results in a smaller net magnetization detected by SQUID-VSM but an increased Kerr rotation detected by MOKE, which is more sensitive towards magnetization changes of the top layers. Note, that in this case, the possible linear range of a GMR sensor is limited by $H_\text{SF}$ instead of $H_\text{C,1}$.\\
In order to confirm our findings and to unravel the underlying reversal processes, micromagnetic simulations were carried out.

\onecolumngrid
\begin{center}
\begin{table}[h!]
  \centering
  \vspace{0.5cm}
  \begin{tabular}{c c c || c c c}
    \toprule
    \toprule
      & \multicolumn{2}{c||}{single p-SAF} & \multicolumn{3}{c}{double p-SAF} \\
    \midrule
      & $L_{\mathrm{bot}}$ & $L_{\mathrm{top}}$ & $L_{\mathrm{bot}}$ & $L_{\mathrm{mid}}$ & $L_{\mathrm{top}}$ \\
    \midrule
    $K_{\mathrm{eff}}$\,(Merg/cm$^3$) & 3.49 & 3.99 & $2.99^{\mathrm{\ref{fig:GMRconfigsSimu}e}}$/\underline{4.49}$^{\mathrm{\ref{fig:GMRconfigsSimu}d,\ref{fig:GMRconfigsSimu}f}}$ & $3.99^{\mathrm{\ref{fig:GMRconfigsSimu}e}}$/\underline{4.49}$^{\mathrm{\ref{fig:GMRconfigsSimu}d,\ref{fig:GMRconfigsSimu}f}}$ & $-0.01$ \\
    $M_{\mathrm{S}}$\,(kemu/cm$^3$) & 1.15 & 1.35 & 1.43 & \underline{1.15}$^{\mathrm{\ref{fig:GMRconfigsSimu}d,\ref{fig:GMRconfigsSimu}e}}$/$1.31^{\mathrm{\ref{fig:GMRconfigsSimu}f}}$ $ $ & $\quad$\underline{1.75}$^{\mathrm{\ref{fig:GMRconfigsSimu}d,\ref{fig:GMRconfigsSimu}e}}$/$1.67^{\mathrm{\ref{fig:GMRconfigsSimu}f}}$ \\
    $A_{\mathrm{ex}}$\,($\mu$erg/cm) & 1.0 & 1.0 & 1.0 & 1.0 & 1.0 \\
    $J_{\mathrm{iex}}$\,(erg/cm$^2$) & \multicolumn{2}{c||}{$-2.5$} & \multicolumn{3}{c}{\hspace{-0.50cm} $-2.5$ \hspace{1.5cm} $-2.5$} \\
    $\alpha$ & 1.0 & 1.0 & 1.0 & 1.0 & 1.0 \\
    \midrule
    $a$\,($\text{\AA}$) & 10 & 10 & 10 & 10 & 10 \\
    $t$\,($\text{\AA}$) & 28 & 29 & 33 & 45 & 10 \\
    \bottomrule
    \bottomrule
  \end{tabular}
   \caption{\small Material parameters of the simulated single p-SAF and double p-SAF systems for all involved layers. $K_{\mathrm{eff}}$ is the effective uniaxial magnetic anisotropy constant, $M_{\mathrm{S}}$ is the saturation magnetization, $A_{\mathrm{ex}}$ is the bulk exchange constant within the layers, $J_{\mathrm{iex}}$ is the interface exchange coupling between the layers, $\alpha$ is the dimensionless damping constant, $a$ is the lateral side length of the simulated nanorod and $t$ is the thickness of the layers. The superscript symbols in the values of $K_{\mathrm{eff}}$ and $M_{\mathrm{S}}$ refer to the used parameters in the subplots of Fig.~\ref{fig:GMRconfigsSimu}. The underlined parameters indicate the basis parameters that are used unless otherwise specified.}
  \label{tab:material}
\end{table}
\end{center}
\twocolumngrid

\section{MICROMAGNETIC SIMULATION}
\pgfplotsset{colormap/Blues-9}
\label{sec:magnum}
\begin{figure*}[tbh!]
\centering
\includegraphics{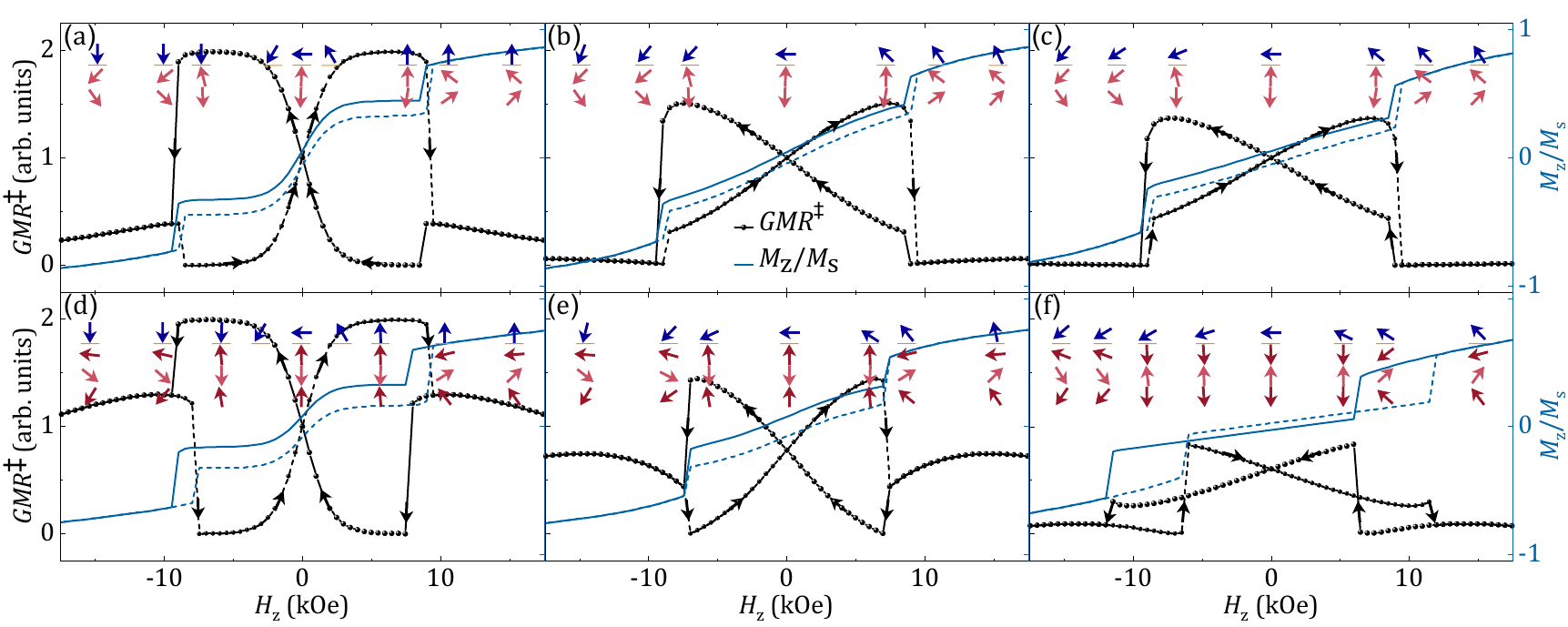}%
 \caption{\small Simulation results for GMR$^{\ddag}$ signals and normalized magnetizations for GMR sensors in out-of-plane fields. GMR sensors with (a-c) single and (d-f) double p-SAF reference system, and free layer (a,c)~$H_{\text{k}}\leq H_{\text{ex}}$, (b,e)~$H_{\text{k}}> H_{\text{ex}}$, and (c,f)~$H_{\text{k}}\gg H_{\text{ex}}$. The arrows illustrate simulated magnetization directions of the respective layers, with blue representing the free layer and red representing the layers of the p-SAF reference system.\label{fig:GMRconfigsSimu}}
\end{figure*}
To understand the experimentally obtained magnetization reversal processes and the corresponding GMR of Fig.~\ref{fig:GMRconfigs}, the finite-element software package magnum.fe~\cite{abert_magnum.fe:_2013} was used to simulate the field dependence of the magnetization of the introduced GMR sensors by means of a spin-chain model. This model consists of a 3D nanorod with a square basal plane of side length $a=10$\,$\text{\AA}$, but with a lateral discretization length much larger than 10\,$\text{\AA}$. This produces a mesh with nodes only along the edges in lateral direction. Along the easy-axis direction (z direction) a fine mesh with a discretization length of approximately 5\,$\text{\AA}$ is used. The reference material parameters for the layers of both investigated p-SAFs are based on the experimental data and are summarized in Table~\ref{tab:material}.

In the micromagnetic simulations single ferromagnetic layers with the properties of the p-SAF multilayers are computed. The modeling is started with a positive external saturation field of 30\,kOe with all layers pointing in the +z direction. The field is applied with an angle of 5$^\circ$ with respect to the easy axis to avoid metastable states. Then, the magnetic field magnitude is decreased stepwise in $-0.5$\,kOe increments to $-30$\,kOe and back to +30\,kOe. After each field-step the micromagnetic state of the system is relaxed for 100\,ns. Note, that the variation of the applied field in the modeling work is performed much faster than that used during the acquisition of the measurement data. However, because a high damping constant ($\alpha=1.0$) is used in the modeling work, a stationary state is obtained within 100\,ns, such that the modeled loops are representative for the experimental loops.

To qualitatively model the GMR~\cite{Dieny1991} from the simulated hysteresis loops, we use the enclosed magnetization angle $\gamma$ of the free layer and the top p-SAF layer per:
\begin{equation}
    \mathrm{GMR^{\ddag}}=1-\cos(\gamma).
\end{equation}
Note, that for simplicity the free layer magnetization is not simulated. Instead it is modeled with a generic $\tanh$ function, where the thickness, the saturation field and the saturation magnetization of the free layer are chosen according to the measured sensor stacks (see Fig.~\ref{fig:GMRconfigs}).

With the spin-chain model and the material parameters of Tab.~\ref{tab:material} we obtain an excellent qualitative agreement with the measurements in both the magnetization data as well as in the GMR, as illustrated in Fig.~\ref{fig:GMRconfigsSimu}. The spin-canting of the p-SAF layers obtained from the simulations also agrees well with that extracted from the GMR measurements of Fig.~\ref{fig:GMRconfigs}. Note, that our simulations also show a small spin-canting of the p-SAF at small fields as hypothesized earlier from Fig.~\ref{fig:extremeWideLR}(a). To achieve this agreement, some, at this point, seemingly arbitrary variations were made to the material parameters of the individual layers of the double p-SAF system. The parameter variations are given by the superscripts, containing the corresponding reference figure to which they refer, in Tab.~\ref{tab:material}.

In the following we will make a more systematic analysis of the influences of the most important parameters for the double p-SAF structures. Based on the reference parameters of Tab.~\ref{tab:material} (underlined values if more than one is given) we start with the saturation magnetization of the layers and vary that of L$_\text{top}$ in the range of 1.19\,kemu/cm$^3$ to 1.95\,kemu/cm$^3$ and that of L$_\text{mid}$ in the range of 0.88\,kemu/cm$^3$ to 1.39\,kemu/cm$^3$, while keeping the properties of L$_\text{bot}$ constant. In both cases a step size of $\Delta M_{\mathrm{S}}=0.02$\,kemu/cm$^3$ is used. 
\begin{figure}
    \centering
    \includegraphics{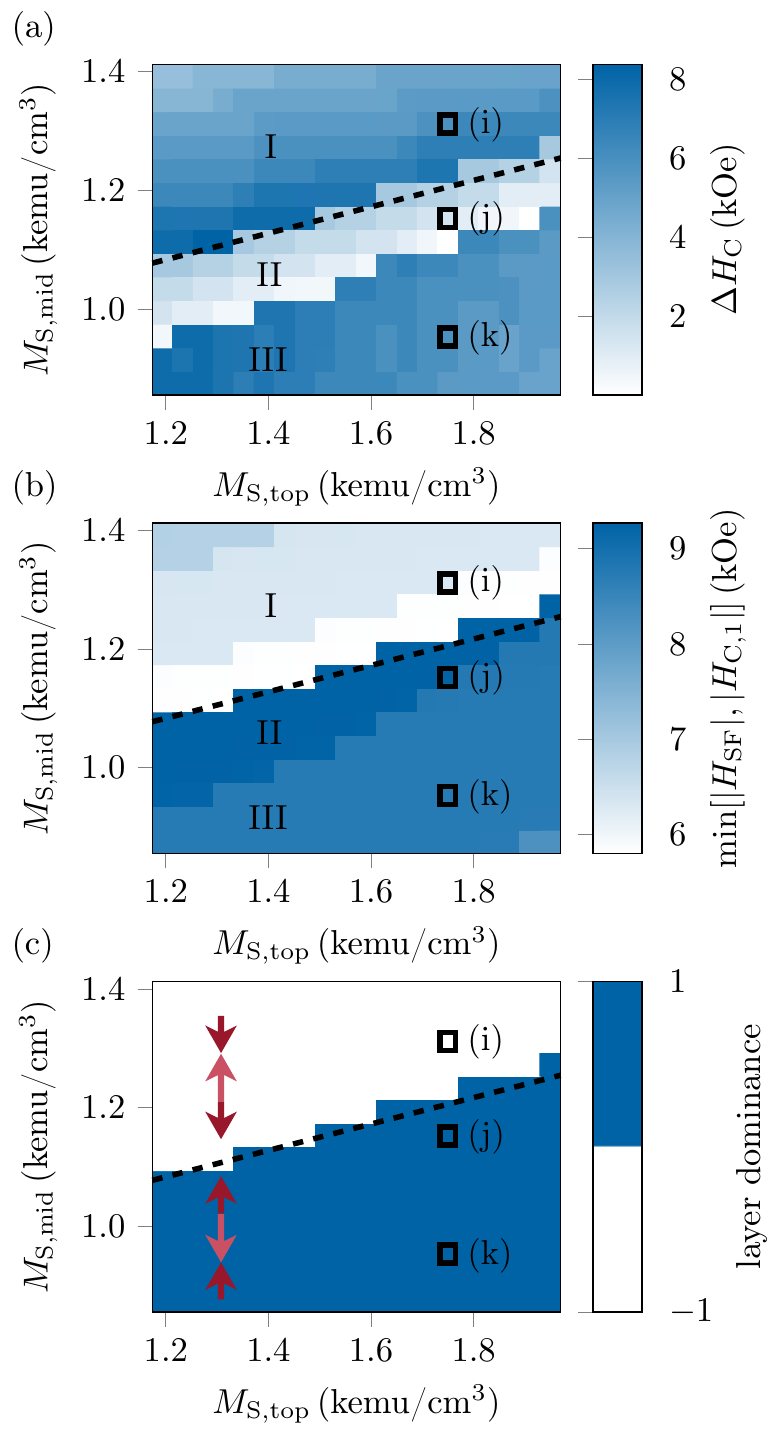}
    \caption{\small Characteristic properties of the hysteresis loops of the double p-SAF system for variations of the saturation magnetizations of the top and the middle p-SAF layer. All material parameters that are not varied are taken from Tab.~\ref{tab:material}. The properties are (a) the hysteresis loop width $\Delta H_{\mathrm{C}}$, (b) the linear range of the sensor stack defined as $\min[|H_{\mathrm{SF}}|,|H_{\mathrm{C,1}}|]$ and (c) the layer dominance with respect to the external field. The arrows illustrate the meaning of the dominance by showing the reversal mechanism of the layers for a decreasing external field coming from positive saturation. The dashed line illustrates the border between the regions of dominance.} 
    \label{fig:ms_phase}
\end{figure}
Figure~\ref{fig:ms_phase} displays three interesting properties of the resulting hysteresis loops, namely the hysteresis width $\Delta H_{\mathrm{C}}$, the linear range, which is defined as $\min[|H_{\mathrm{C,1}}|,|H_{\mathrm{SF}}|]$ (see Fig.~\ref{fig:pSAFs-SpinFlip}), and the layer dominance of the p-SAF with respect to the external field. As shown in the inset of Fig.~\ref{fig:ms_phase}(c), a value of $-1$ means that, when coming from saturation, the top layer switches into an antiparallel direction with respect to the external magnetic field, while a value of $+1$ means that the middle layer switches, which leaves the top layer magnetization direction and external field in a parallel configuration.
Obviously, the two regions show either the middle p-SAF layer or the outer two ones being aligned with the external field. 
In contrast to Figs.~\ref{fig:ms_phase}(b) and (c) three regions appear (labeled with I, II and III) in the minor loop hysteresis width of Fig.~\ref{fig:ms_phase}(a). While the border between regions I and II is identical to that of Fig.~\ref{fig:ms_phase}(c), it is not a priori clear why there is an additional border between regions II and III. 

\begin{figure}
    \centering
    \begin{adjustwidth}{-0.5cm}{-0.5cm}
    \includegraphics{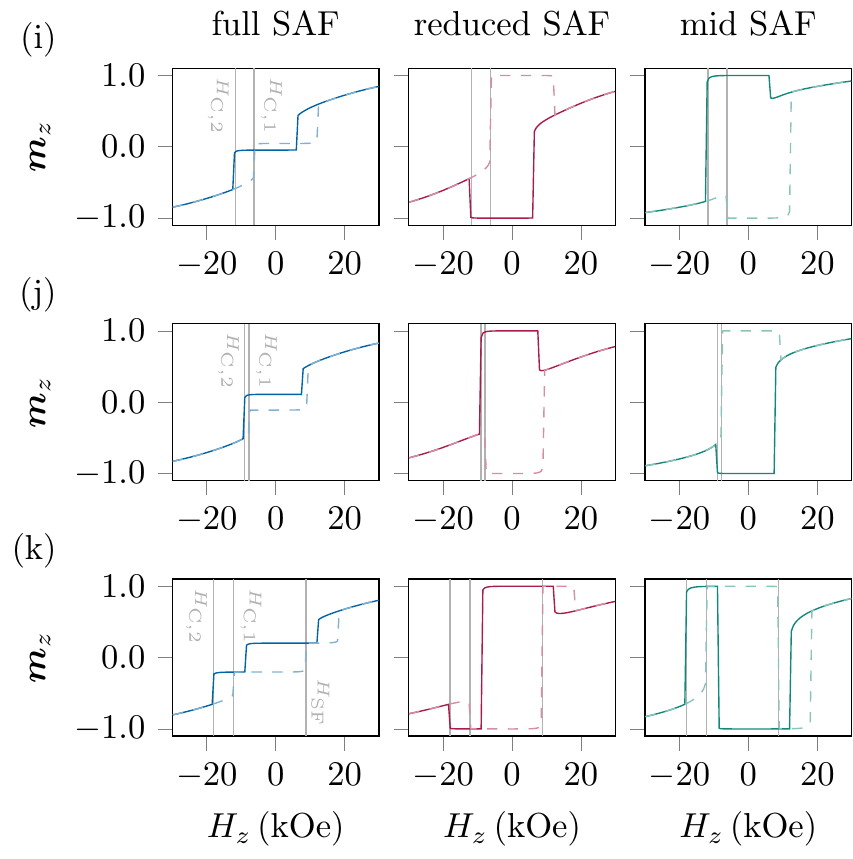}
    \caption{\small   Normalized z component of the magnetization of the full double p-SAF system, the middle p-SAF layer and a reduced p-SAF layer as introduced in Eq.~\ref{eq:reduced}. The magnetization for decreasing fields coming from positive saturation is illustrated with a solid line and that of for increasing fields coming from negative saturation is shown with a dashed line. (i), (j) and (k) refer to the phase points marked in Fig.~\ref{fig:ms_phase}. The vertical lines indicate the switching fields $H_{\mathrm{C,1}}$ and $H_{\mathrm{C,2}}$ and the spin-flip field $H_{\mathrm{SF}}$ (introduced in Fig.~\ref{fig:pSAFs-SpinFlip}).}
    \label{fig:loops}
  \end{adjustwidth}
\end{figure}
To gain deeper insights into the magnetization dynamics, we investigate three phase points of Fig.~\ref{fig:ms_phase} in more detail. Figure~\ref{fig:loops} illustrates the hysteresis loops of phase points (i), (j) and (k). In detail, the normalized magnetization of the full double p-SAF is shown in the first column and that of L$_\text{mid}$ is shown in the last column. To lower the complexity of the system L$_\text{top}$ and L$_\text{bot}$ are reduced to a single layer with the following properties:
\begin{eqnarray}
\label{eq:reduced}
 K_{\mathrm{eff,red}}&=&\frac{K_{\mathrm{eff,top}}t_{\mathrm{top}}+K_{\mathrm{eff,bot}}t_{\mathrm{bot}}}{t_{\mathrm{top}}+t_{\mathrm{bot}}}\nonumber\\
 M_{\mathrm{S,red}}&=&\frac{M_{\mathrm{S,top}}t_{\mathrm{top}}+M_{\mathrm{S,bot}}t_{\mathrm{bot}}}{t_{\mathrm{top}}+t_{\mathrm{bot}}}
\end{eqnarray}
This reduction implicitly assumes a parallel alignment of L$_\text{top}$ and L$_\text{bot}$, which is not strictly true at least during the reversal. But this simplified picture is much more instructive and most importantly it is sufficient to characterize the magnetization dynamics correctly. In the middle column of Fig.~\ref{fig:loops} the dynamics of this reduced layer is displayed. Additionally, the magnetizations for decreasing fields from positive to negative saturation are displayed with solid lines, while the magnetizations for increasing fields from negative to positive saturation are displayed with dashed lines. 

A typical hysteresis loop for region I in Fig.~\ref{fig:ms_phase} with high $M_{\mathrm{S,mid}}$ and low $M_{\mathrm{S,top}}$, and thus low $M_{\mathrm{S,red}}$ is shown in Fig.~\ref{fig:loops}(i). In agreement with Fig.~\ref{fig:ms_phase}(c), we see that L$_\text{mid}$ dominates the magnetization process. Hence, the switching field of L$_\text{mid}$ determines the outer part of the hysteresis loop (vertical line with label $H_{\mathrm{C},2}$ in Fig.~\ref{fig:loops}) and the field at which the reduced p-SAF layer switches back to an antiparallel state determines the inner part of the hysteresis loop (vertical line with label $H_{\mathrm{C},1}$ in Fig.~\ref{fig:loops}). We will call the latter field backswitching field. Since the switching fields are inversely proportional to the saturation magnetization, we see an increase of $H_{\mathrm{C,1}}$ towards lower values of $M_{\mathrm{S,top}}$ for fixed $M_{\mathrm{S,mid}}$ in Fig.~\ref{fig:ms_phase}(a). This results in an increasing linear range and a decrease of $\Delta H_{\mathrm{C}}$.
\begin{figure}
    \centering
    \begin{adjustwidth}{-0.5cm}{-0.5cm}
    \includegraphics{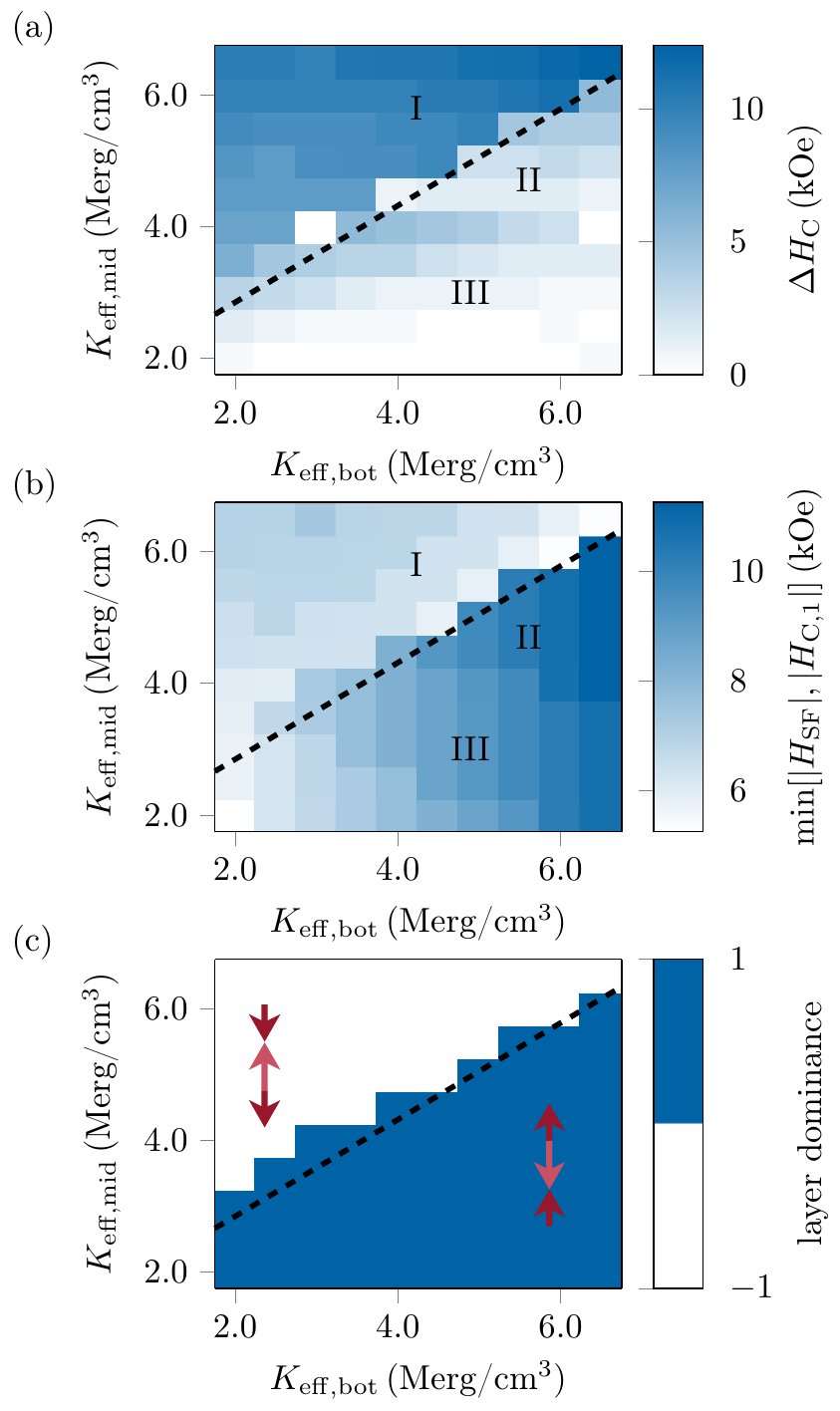}
    \caption{\small   Characteristic properties of the hysteresis loops of the double p-SAF system for variations of the effective uniaxial magnetic anisotropy of the top and the middle p-SAF layer. All material parameters that are not varied are taken from Tab.~\ref{tab:material}. The properties are (a) the hysteresis loop width $\Delta H_{\mathrm{C}}$, (b) the linear range of the sensor stack defined as $\min[|H_{\mathrm{SF}}|,|H_{\mathrm{C,1}}|]$ and (c) the layer dominance with respect to the external field. The arrows illustrate the meaning of dominance by showing the reversal mechanism of the layers for a decreasing external field coming from positive saturation. The dashed line illustrates the border between the regions of dominance.} 
    \label{fig:K_phase}
    \end{adjustwidth}
\end{figure}
If $M_{\mathrm{S,mid}}$ decreases from 1.31\,kemu/cm$^3$ to 1.11\,kemu/cm$^3$ and the saturation magnetization of the reduced layer remains fixed, we arrive at phase point (j) and the hysteresis loops of Fig.~\ref{fig:loops}(j). Here, the reduced p-SAF layer is dominating the reversal. Since it has a higher magnetic moment its switching field $H_{\mathrm{C},2}$ is much lower than that of L$_\text{mid}$ in region I. It is slightly higher than the backswitching field $H_{\mathrm{C},1}$ of L$_\text{mid}$ resulting in a very small $\Delta H_{\mathrm{C}}$. Due to the change of dominance $H_{\mathrm{C},1}$ is now determined by L$_\text{mid}$. Hence, the linear range in region II is significantly higher than that in region I. The effect of the decrease of $M_{\mathrm{S,mid}}$ from region I to region II can also be clearly seen in the SQUID-VSM measurements of Fig.~\ref{fig:pSAFs-Switching}, where the decrease is caused by changing the layer numbers in L$_{\mathrm{mid}}$. Within region II the linear range further increases with decreasing $M_{\mathrm{S,mid}}$. In contrast, $\Delta H_{\mathrm{C}}$ decreases with decreasing $M_{\mathrm{S,mid}}$, since $H_{\mathrm{C},1}$ comes closer to $H_{\mathrm{C},2}$. Note, due to the antiferromagnetic exchange coupling, the magnitude of the switching field from an antiparallel to a parallel state is higher than that back to the antiparallel state. 
The experimental hysteresis loops for different layer numbers in L$_{\mathrm{mid}}$ shown in Fig.~\ref{fig:pSAFs-Switching} agree very well with the spin-chain simulations, and thus confirm the influence of an increasing magnetic moment of the middle p-SAF layer in region II.

For even lower $M_{\mathrm{S,mid}}$ we see an abrupt increase of $\Delta H_{\mathrm{C}}$ in Fig.~\ref{fig:ms_phase}(a). The reason is illustrated in Fig.~\ref{fig:loops}(k). At the switching field of the dominating reduced p-SAF layer a spin flip occurs. The reason is that the switching field of L$_\text{mid}$ is much larger than that of the reduced layer. Therefore the reduced layer becomes dominant after the spin flip and determines both the switching field $H_{\mathrm{C},2}$ and the backswitching field $H_{\mathrm{C},1}$. 
This switching behavior can also clearly be seen in Fig.~\ref{fig:pSAFs-SpinFlip}.

But not only the saturation magnetization has an influence on the hysteresis of the double p-SAF. However, since a systematic analysis of all possible influences goes beyond the scope of this work, we will exemplarily discuss the variation of the effective magnetic anisotropies of the bottom and the middle layer with constant $K_{\mathrm{eff,top}}=-0.01$\,Merg/cm$^3$ in the following.

Figure~\ref{fig:K_phase} displays the hysteresis loop width, the linear range and the layer dominance for $K_{\mathrm{eff}}$ of both layers in the range of 1.99\,Merg/cm$^3$ to 6.49\,Merg/cm$^3$ with $\Delta K_{\mathrm{eff}}=0.5$\,Merg/cm$^3$. We again see three regions that represent the same types of hysteresis loops as those in Fig.~\ref{fig:ms_phase}. In region I, L$_\text{mid}$ dominates the reversal process under an applied magnetic field, while in regions II and III the reduced p-SAF layer dominates. Region II again shows spin-flip processes. 
Figure~\ref{fig:K_phase} again proves that the magnetic anisotropy is as important as the saturation magnetization of the individual layers if we want to describe the reversal mechanism of the double p-SAF. 


\section{CONCLUSION}
In conclusion, we have developed p-SAFs with up to 10\,kOe exchange fields based on Co/Pt multilayers, using both single and double structures for interlayer exchange coupling. These p-SAFs were implemented in GMR sensors with perpendicular reference layer and in-plane free layer, yielding up to 8\,kOe dynamic field range. We designed different free layer systems, further utilizing Co/Pt multilayers, which yield varied magnetic anisotropies in out-of-plane fields. In combination with our p-SAF structures, the magnetic anisotropy variation revealed spin-canting effects in the interlayer exchange coupled reference system, predominantly in the double p-SAF GMR sensors. Micromagnetic simulations based on finite-element spin-chain models further investigated spin-canting effects and p-SAF switching behavior in respect to individual layer saturation magnetization and magnetic anisotropy. Experimental data and simulation results highly agree with one another. The results provide a better understanding of magnetoresistive sensor design and design potentials. However, while double p-SAFs offer a very large design flexibility, one needs to be very careful that uncontrollable variations in the material properties due to the growth of the structure, do not lead to undesired effects.

\section{Acknowledgments}
This study has been supported by the Austrian Science Fund under Grant I2214-N20, and is gratefully acknowledged. The support of the CD-Labor AMSEN (financed by the Federal Ministry of Economics, Family and Youth, the National Foundation for Research, Technology and Development) is acknowledged.


%

\end{document}